\documentclass[%
 reprint,
superscriptaddress,
 amsmath,amssymb,
 aps,
prb,
]{revtex4-2}
\usepackage{subscript}

\usepackage{graphicx}
\usepackage{dcolumn}
\usepackage[utf8]{inputenc}
\usepackage{newunicodechar}
\newunicodechar{₂}{$_2$}
\usepackage{bm}
\usepackage{hyperref}
\usepackage{soul}

\usepackage{graphicx}
\usepackage{dcolumn}
\usepackage{bm}
\usepackage{amsmath}
\usepackage{amssymb}
\usepackage{color}
\usepackage{hyperref}
\usepackage{siunitx}

\begin{document}


\title{A GPU-Native Atom-Resolved Framework for
Moir\'e Analysis of Twisted Bilayer Graphene}

\author{Seongin Kim}
\affiliation{ALL.AI Inc., Seongnam-si, Gyeonggi-do 13122, Korea}

\author{Yong-Gak Kim}
\affiliation{ALL.AI Inc., Seongnam-si, Gyeonggi-do 13122, Korea}

\author{Sungho Choi}
\affiliation{ALL.AI Inc., Seongnam-si, Gyeonggi-do 13122, Korea}

\author{Jinyeong Wang}
\email{jinyeong.wang@allai.co.kr}
\affiliation{ALL.AI Inc., Seongnam-si, Gyeonggi-do 13122, Korea}

\author{Sangyun Lee}
\email{sangyun.lee@ufl.edu}
\affiliation{National High Magnetic Field Laboratory High B/T Facility and Department of Physics,
University of Florida, Gainesville, Florida 32611, USA}

\date{\today}

\begin{abstract}

Atom-resolved modeling of twisted bilayer graphene (TBG) becomes
increasingly demanding as the twist angle decreases and the
characteristic moir\'e length scale grows. We develop a GPU-native
computational framework for atom-resolved structural analysis of
rigid TBG using NVIDIA Warp, coupled to NVIDIA Omniverse for
interactive visualization.
Atomic coordinates are generated and updated directly on the GPU,
and local interlayer displacements are reduced modulo the graphene
primitive lattice to classify AA-, AB-, and BA-like stacking
environments. Bond-connected AA-like sites are grouped into
individual domains, from which representative AA centers are
identified. The moir\'e length is characterized using the minimum
and mean distances, $L_{\min}$ and $L_{\mathrm{ave}}$, between a
representative central AA domain and its six lattice-neighbor AA~domains.
Over the investigated twist-angle range
$0.5^\circ \leq \theta \leq 3.0^\circ$, the calculated moir\'e
lengths follow the analytical geometric relation
$L_{\mathrm{m}}=a/[2\sin(\theta/2)]$, with maximum absolute
relative errors of 5.00\% for $L_{\min}$ and 2.06\% for
$L_{\mathrm{ave}}$.
GPU performance is evaluated using an identical NVIDIA Warp
implementation on CPU and GPU backends. For a system containing
1,778,700 atoms in both layers, the core calculation latency decreases from
52.718~ms on a single-threaded Warp CPU backend to 0.2706~ms on an
NVIDIA RTX 5090 GPU, corresponding to a speedup of 194.8$\times$.
The benchmark excludes initialization, heatmap generation,
user-interface updates, and rendering.
The framework provides a scalable structural-analysis layer for
large atom-resolved moir\'e systems and establishes a basis for
future coupling to electronic-structure calculations.

\end{abstract}


\maketitle

\section{\label{sec:Introduction}Introduction}

Twisted bilayer graphene (TBG) is a prototypical moir\'e material in
which a small relative rotation between two graphene layers generates
a long-wavelength superlattice and strongly modifies the electronic
structure \cite{Bistritzer2011}.
Near the first magic angle, narrow electronic bands enhance interaction
effects and give rise to correlated insulating behavior,
superconductivity, and other strongly correlated phenomena
\cite{Cao2018Insulator,Cao2018,Kerelsky2019}.
As the twist angle decreases,
the moir\'e period increases rapidly, leading to large real-space
supercells and a corresponding growth in the number of atomic degrees
of freedom. Explicit atom-resolved calculations therefore become
increasingly demanding, particularly when strain, lattice relaxation,
structural disorder, or local defects are considered.
For atom-resolved models spanning comparable moir\'e-scale regions,
this rapid increase in moir\'e length requires progressively larger
atomic systems as the twist angle decreases.

A broad range of computational approaches has been developed for TBG.
Continuum models efficiently describe the low-energy moir\'e electronic
structure \cite{Bistritzer2011,Carr2019}, whereas atomistic and
real-space methods retain microscopic information on local stacking,
lattice reconstruction, strain, and defects
\cite{Nam2017,Dey2023,Baldo2023}. 
Several software frameworks have been developed for computational studies of moir\'e materials. TBPLaS provides efficient real-space tight-binding calculations~\cite{Li2023TBPLaS}, while Twister focuses on the construction of commensurate moir\'e superlattices and structural relaxation through an interface to classical force-field calculations~\cite{Naik2022Twister}. 
More recently, MoireStudio has integrated commensurate-structure generation, relaxation, tight-binding and continuum-model construction, parallel computation, and visualization within a general framework for twisted materials~\cite{Yu2026MoireStudio}. 
NanoStudio provides automated atomistic construction of graphene-based and other honeycomb-lattice nanostructures, including twisted graphene with arbitrary commensurate twist angles~\cite{Qin2026NanoStudio}. 
Interactive tools such as PyAtoms provide a complementary approach for rapidly exploring moir\'e structures using a Fourier-space representation of ideal atomic lattices~\cite{Gutierrez2024PyAtoms}.

Despite this progress, existing packages primarily emphasize commensurate-supercell construction, structural relaxation, electronic-structure modeling, or interactive visualization. 
Comparatively less attention has been given to a GPU-resident workflow that directly couples explicit atom-resolved coordinate updates to local stacking classification, connected AA-domain identification, quantitative moir\'e-length extraction, and interactive three-dimensional visualization. 
In particular, the present framework targets repeated angle-dependent structural analysis while minimizing intermediate CPU--GPU data transfers and avoiding a separate Euclidean neighbor search for the local AA-domain lattice.
Unlike workflows in which visualization is applied as a separate post-processing stage, the present framework couples GPU-resident structural updates and registry analysis directly to the interactive visualization loop.
In a
conventional atomistic workflow, structural parameters are typically
specified before the calculation, followed by structure generation,
analysis, and subsequent visualization. For interactive exploration,
however, a change in a control parameter such as the twist angle
should ideally update the atomic coordinates, local stacking information, quantitative structural descriptors, and visual representation
within the same computational loop. Achieving this efficiently
requires minimizing unnecessary data movement between CPU-based
processing, GPU computation, and visualization.

Here, we develop a GPU-native atom-resolved framework for structural analysis of rigid TBG using NVIDIA Warp~\cite{Macklin2022Warp}. 
The method evaluates local
AA-, AB-, and BA-like stacking directly from GPU-generated atomic
coordinates and identifies representative AA domains for quantitative
moir\'e characterization.

The principal distinction of the present framework is the integration of explicit atom-resolved geometry, GPU-resident registry analysis, and interactive visualization within a unified computational pipeline, rather than the introduction of a new physical model of TBG.

For the ideal rigid structures studied here, the moir\'e length is characterized from the six lattice-neighbor AA domains surrounding a representative central AA domain. The extracted structural descriptors are validated against the analytical moir\'e period over a range of twist angles. We further evaluate the computational performance using identical NVIDIA Warp implementations on CPU and GPU backends, while NVIDIA Omniverse provides an interactive visualization layer for the resulting atom-resolved structures and moir\'e descriptors.

\section{Computational methodology}

\subsection{GPU construction of atom-resolved graphene layers}

Each graphene layer is represented as a finite hexagonal lattice whose
size is controlled by an integer shell parameter $s$. Defining

\begin{equation}
n=s+1,
\end{equation}

the number of carbon atoms in a single layer is

\begin{equation}
N_{\mathrm{atom}}=6n^2,
\label{eq:natom}
\end{equation}

while the number of nearest-neighbor bonds is

\begin{equation}
N_{\mathrm{bond}}
=
3n(3n-1).
\label{eq:nbond}
\end{equation}

Atomic coordinates are generated directly on the GPU. Each GPU thread
is assigned to one carbon site and determines its position from the
corresponding row and column indices of the finite hexagonal lattice.
The resulting coordinates are stored in an immutable reference array,
$\mathbf{r}_i^0$, which is retained throughout subsequent twist-angle
updates.

Unless otherwise specified, $N_{\mathrm{atom}}$ denotes the number
of atoms in a single graphene layer, whereas the performance
benchmarks use the total bilayer atom count

\begin{equation}
N_{\mathrm{total}}=2N_{\mathrm{atom}}.
\end{equation}

In the coordinate convention used in the implementation, the
primitive graphene lattice vectors are

\begin{equation}
\mathbf{a}_1
=
(\sqrt{3}d_{\mathrm{CC}},0),
\end{equation}

and

\begin{equation}
\mathbf{a}_2
=
\left(
\frac{\sqrt{3}d_{\mathrm{CC}}}{2},
\frac{3d_{\mathrm{CC}}}{2}
\right),
\end{equation}

where \(d_{\mathrm{CC}}=1.42\) Å is the nearest-neighbor carbon--carbon distance. 
The corresponding graphene lattice constant is

\begin{equation}
a=\sqrt{3}d_{\mathrm{CC}}.
\end{equation}

These primitive vectors and the resulting finite hexagonal topology
define the immutable reference configuration used throughout the
computational pipeline.

\subsection{Reference bilayer configuration}

The generated graphene coordinates are stored in an immutable
reference array, $\mathbf{r}_i^0=(x_i^0,y_i^0,0)$.
At zero twist angle, the bottom and top layers are initialized as

\begin{equation}
\mathbf{r}^{\mathrm{bottom}}_i
=
(x_i^0,y_i^0,0),
\end{equation}

and

\begin{equation}
\mathbf{r}^{\mathrm{top}}_i
=
(x_i^0,y_i^0,d_{\mathrm{inter}}),
\end{equation}

where $d_{\mathrm{inter}}=3.35$~\AA\ is the prescribed
interlayer separation.
The reference coordinates $\mathbf{r}_i^0$ are not overwritten during
the simulation. For each requested twist angle, the top-layer
coordinates are regenerated directly from this reference
configuration rather than from the previously rotated configuration.
This prevents the accumulation of numerical drift during continuous
angle sweeps.


\subsection{GPU rotation of the top layer}

The center of the finite hexagonal graphene layer is chosen as the
in-plane rotation origin. For a user-defined twist angle $\theta$,
the top-layer coordinates are updated according to

\begin{equation}
\begin{pmatrix}
x_i^{\mathrm{top}}\\
y_i^{\mathrm{top}}
\end{pmatrix}
=
\begin{pmatrix}
\cos\theta & -\sin\theta\\
\sin\theta & \cos\theta
\end{pmatrix}
\begin{pmatrix}
x_i^0\\
y_i^0
\end{pmatrix}.
\label{eq:rotation}
\end{equation}

The transformation is evaluated independently for each atomic site
using one GPU thread per atom, while the bottom layer remains fixed.


\subsection{Atomic-site displacement and local registry}

For each atomic site $i$, the raw in-plane displacement between the
corresponding top- and bottom-layer reference positions is calculated
as

\begin{equation}
\mathbf{u}^{\mathrm{raw}}_i
=
\mathbf{r}^{\mathrm{top}}_{i,xy}
-
\mathbf{r}^{\mathrm{bottom}}_{i,xy}.
\label{eq:raw_u}
\end{equation}

For a rigid twist this is equivalent to

\begin{equation}
\mathbf{u}^{\mathrm{raw}}_i
=
[R(\theta)-I]\mathbf{r}_{i,xy}^{0}.
\end{equation}

Here, the index $i$ denotes a fixed carbon site in the zero-angle
reference configuration; nearest-neighbor atoms are not rematched as
the twist angle is changed.


\subsection{Periodic reduction of the displacement field}

Local stacking is invariant under integer translations of the graphene primitive lattice. Therefore, the raw displacement is first
represented in the primitive-lattice basis,

\begin{equation}
\mathbf{u}^{\mathrm{raw}}_i
=
c_{1,i}\mathbf{a}_1+
c_{2,i}\mathbf{a}_2.
\end{equation}

For the lattice convention adopted here,

\begin{equation}
c_{2,i}
=
\frac{u^{\mathrm{raw}}_{i,y}}
{3d_{\mathrm{CC}}/2},
\end{equation}

and

\begin{equation}
c_{1,i}
=
\frac{u^{\mathrm{raw}}_{i,x}}
{\sqrt{3}d_{\mathrm{CC}}}
-\frac{1}{2}c_{2,i}.
\end{equation}

Integer lattice translations are removed using

\begin{equation}
\widetilde{c}_{\alpha,i}
=
c_{\alpha,i}-\lfloor c_{\alpha,i}\rfloor,
\qquad \alpha=1,2,
\end{equation}

and the reduced displacement is reconstructed as

\begin{equation}
\mathbf{u}^{\mathrm{local}}_i
=
\widetilde{c}_{1,i}\mathbf{a}_1+
\widetilde{c}_{2,i}\mathbf{a}_2.
\label{eq:local_displacement}
\end{equation}


\subsection{AA, AB, and BA registry classification}

In the coordinate convention used in the present implementation, the
three high-symmetry registry configurations are represented by

\begin{equation}
\mathbf{u}_{\mathrm{AA}}=(0,0),
\end{equation}

\begin{equation}
\mathbf{u}_{\mathrm{AB}}=(0,d_{\mathrm{CC}}),
\end{equation}

and

\begin{equation}
\mathbf{u}_{\mathrm{BA}}=(0,-d_{\mathrm{CC}}).
\end{equation}

Because the reduced displacement $\mathbf{u}_i^{\mathrm{local}}$ is restricted to the primitive displacement cell spanned by $\mathbf{a}_1$ and $\mathbf{a}_2$, the periodic-image search can be restricted to the four neighboring images relevant to each high-symmetry registry. In the primitive-lattice basis, the AB and BA reference displacements are

\begin{equation}
\mathbf{u}_{\mathrm{AB}}
=
-\frac{1}{3}\mathbf{a}_1
+
\frac{2}{3}\mathbf{a}_2,
\end{equation}

and

\begin{equation}
\mathbf{u}_{\mathrm{BA}}
=
\frac{1}{3}\mathbf{a}_1
-
\frac{2}{3}\mathbf{a}_2,
\end{equation}

respectively.

The periodic registry distance is therefore evaluated as

\begin{equation}
d_X(i)
=
\min_{(m,n)\in\mathcal{P}_X}
\left|
\mathbf{u}_i^{\mathrm{local}}
-
\mathbf{u}_X
-
m\mathbf{a}_1
-
n\mathbf{a}_2
\right|,
\end{equation}

where $X\in\{\mathrm{AA},\mathrm{AB},\mathrm{BA}\}$, with the registry-dependent image sets

\begin{equation}
\mathcal{P}_{\mathrm{AA}}
=
\{0,1\}\times\{0,1\},
\end{equation}

\begin{equation}
\mathcal{P}_{\mathrm{AB}}
=
\{0,1\}\times\{-1,0\},
\end{equation}

and

\begin{equation}
\mathcal{P}_{\mathrm{BA}}
=
\{0,1\}\times\{1,2\}.
\end{equation}

The local stacking label is then assigned according to

\begin{equation}
X_i
=
\arg\min_X d_X(i).
\end{equation}

\subsection{AA-domain identification and center detection}

The discrete registry classification generally assigns multiple
neighboring atomic sites to the same AA-like region. These sites
should not be treated as independent moir\'e centers. We therefore
group bond-connected AA-classified sites into spatially connected
AA domains using the nearest-neighbor bond topology of the graphene
lattice.

The connected components are identified using a GPU-based union--find procedure. 
Two AA-classified sites belong to the same domain when they are connected through the nearest-neighbor graphene bond network.

The precomputed nearest-neighbor bond topology is reused for the GPU-based union--find construction, avoiding a separate spatial neighbor search. Atomic compare-and-swap operations are used for concurrent unions, followed by parent compression.

For each connected AA domain $\mathcal{C}_{\alpha}$, a representative
center is defined as the site having the minimum AA-registry distance,

\begin{equation}
i_{\alpha}^{\mathrm{AA}}
=
\underset{i\in\mathcal{C}_{\alpha}}{\mathrm{argmin}}\,
d_{\mathrm{AA}}(i).
\label{eq:aa_center}
\end{equation}

The corresponding AA-center position is

\begin{equation}
\mathbf{R}_{\alpha}^{\mathrm{AA}}
=
\mathbf{r}_{i_{\alpha}^{\mathrm{AA}}}.
\label{eq:aa_center_position}
\end{equation}

This procedure assigns a single representative center to each connected AA-like region and prevents neighboring AA-classified atomic sites belonging to the same physical domain from being interpreted as independent moir\'e centers.

\subsection{Representative moir\'e-length extraction}

For the ideal rigid TBG structures considered here, the AA domain closest to the geometric center of the finite graphene flake is selected as the reference domain. The AA-domain centers are assigned integer coordinates on the triangular moir\'e lattice, and the six logical nearest-neighbor domains of the reference center are identified from the corresponding lattice-index table. These neighbors correspond to the six adjacent sites of the triangular AA-domain lattice rather than being selected by a separate Euclidean nearest-neighbor search.

For each AA-domain center, the displacement \((\Delta x,\Delta y)\) relative to the reference center is projected onto the moir\'e-lattice frame according to
\[
u=\cos(\theta/2)\Delta x-\sin(\theta/2)\Delta y,
\]
\[
v=-\sin(\theta/2)\Delta x-\cos(\theta/2)\Delta y.
\]
Using \(L_{\mathrm{theory}}\) as the rigid-bilayer moir\'e period, the column index is assigned from the nearest integer multiple of \((\sqrt{3}/2)L_{\mathrm{theory}}\), while the triangular lattice is represented using staggered half-row indices constrained by the column parity. The six logical neighbors correspond to the lattice offsets
\[
(0,\pm2),\qquad(\pm1,\pm1).
\]

Only valid, unambiguous AA-domain centers are used. To suppress finite-flake boundary artifacts, AA-domain centers lying within a margin of $2d_{\mathrm{CC}}$ from the hexagonal sample boundary are excluded from the quantitative moir\'e-length analysis.

The six corresponding center-to-center distances are
\begin{equation}
L_k =
\left|
\mathbf{R}^{\mathrm{AA}}_k
-
\mathbf{R}^{\mathrm{AA}}_0
\right|,
\qquad k=1,\ldots,6.
\end{equation}

Two structural descriptors are extracted from these distances.
The minimum local moir\'e length is

\begin{equation}
L_{\min}
=
\min_{k=1,\ldots,6} L_k,
\end{equation}

while the mean nearest-neighbor moir\'e length is

\begin{equation}
L_{\mathrm{ave}}
=
\frac{1}{6}
\sum_{k=1}^{6}L_k.
\end{equation}

For comparison, the geometrical moir\'e period for an ideal rigid
bilayer is

\begin{equation}
L_{\mathrm{theory}}
=
\frac{a}{2\sin(\theta/2)},
\qquad
a=\sqrt{3}d_{\mathrm{CC}}.
\end{equation}

The corresponding absolute relative errors are defined as

\begin{equation}
\epsilon_{\min}
=
100
\frac{
|L_{\min}-L_{\mathrm{theory}}|
}{
L_{\mathrm{theory}}
},
\end{equation}

and

\begin{equation}
\epsilon_{\mathrm{ave}}
=
100
\frac{
|L_{\mathrm{ave}}-L_{\mathrm{theory}}|
}{
L_{\mathrm{theory}}
}.
\end{equation}

Among these two descriptors, $L_{\mathrm{ave}}$ provides the primary estimate of the moir\'e length, whereas $L_{\min}$ is retained as a local minimum-distance descriptor that is more sensitive to atomic-site discretization.


\subsection{GPU implementation and interactive visualization}

The coordinate transformations and atom-resolved registry calculations are implemented using NVIDIA Warp \cite{Macklin2022Warp} as parallel GPU operations. Atomic
sites are processed independently wherever possible, allowing
$\mathbf{u}^{\mathrm{raw}}_i$,
$\mathbf{u}^{\mathrm{local}}_i$,
$d_{\mathrm{AA}}(i)$,
$d_{\mathrm{AB}}(i)$, and
$d_{\mathrm{BA}}(i)$
to be evaluated concurrently over the simulated structure.

For interactive inspection of the local registry, the atom-resolved
$d_{\mathrm{AA}}$ values are additionally projected onto a regular
two-dimensional analysis grid. Each atomic-site value contributes to
nearby grid points through a Gaussian spatial kernel, and the
continuous field is evaluated as

\begin{equation}
d_{\mathrm{AA}}^{\mathrm{grid}}(\mathbf{r})
=
\frac{
\sum_i
d_{\mathrm{AA}}(i)
w_i(\mathbf{r})
}{
\sum_i
w_i(\mathbf{r})
},
\label{eq:daa_grid}
\end{equation}

where \(w_i(\mathbf r)\) is the Gaussian weight associated with atomic site \(i\),

\begin{equation}
w_i(\mathbf r)
=
\exp\left[
-\frac{|\mathbf r-\mathbf r_i|^2}{2\sigma_g^2}
\right].
\label{eq:w_i}
\end{equation}

The continuous registry field is evaluated on a $1024\times1024$ analysis grid using a Gaussian kernel with smoothing width
$\sigma_g=0.5d_{\mathrm{CC}}=0.71$~\AA.
Contributions are truncated beyond a cutoff radius of
$d_{\mathrm{CC}}=1.42$~\AA.
The resulting gridded field is used only for visualization and does not enter the AA-domain identification or quantitative moir\'e-length analysis.

For CUDA execution, the resulting RGBA analysis image is passed directly from the device buffer to the Omniverse image provider without CPU readback. This maintains a GPU-resident path for the registry-heatmap visualization, while the quantitative structural descriptors remain derived from the underlying atom-resolved registry data.

The resulting field is displayed in a dedicated moir\'e-analysis
view together with the detected AA centers and the connections used
to evaluate the local moir\'e lengths.

The computational workflow is organized to retain atomic-coordinate
and registry data on the GPU wherever permitted by downstream
operations, thereby reducing unnecessary CPU--GPU data transfers.
The computed atomic coordinates and structural descriptors are
coupled to NVIDIA Omniverse for interactive three-dimensional
visualization of the bilayer structure and the corresponding
moir\'e registry.
\begin{figure*}
\centering
\includegraphics[width=2\columnwidth]{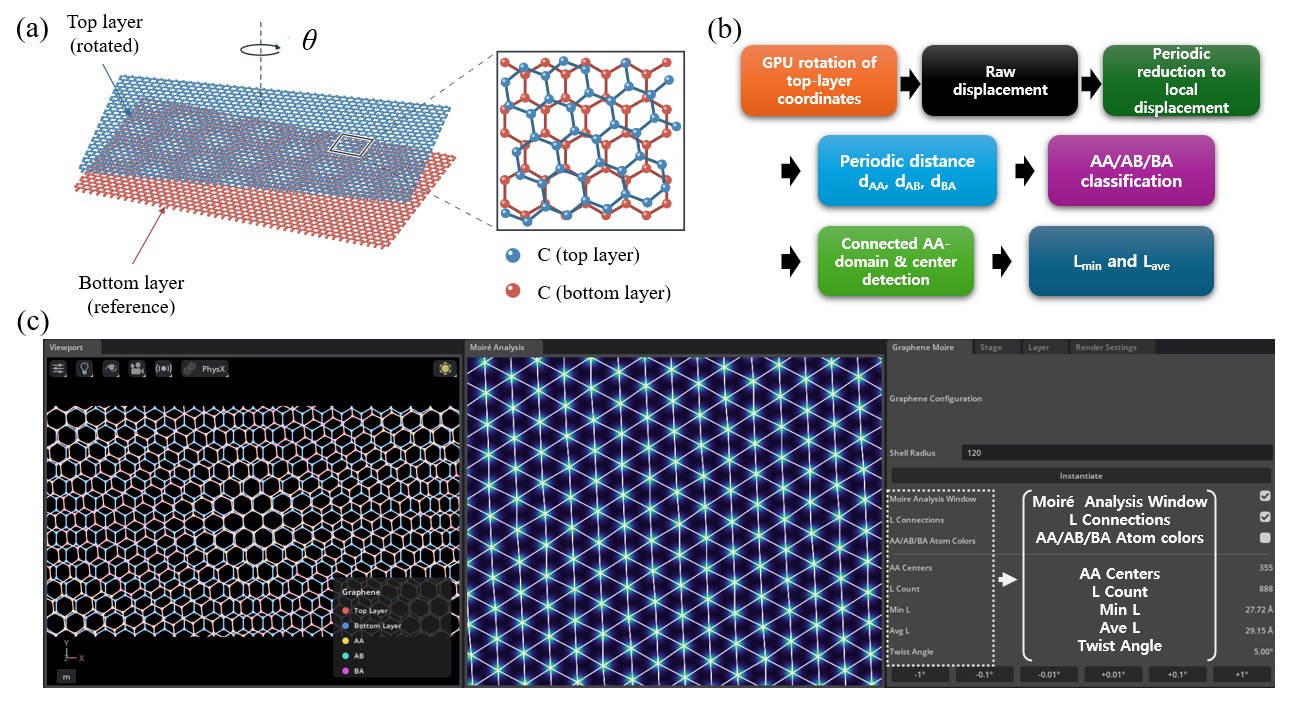}
\caption{\label{fig:rt} 
Schematic overview of the GPU-native computational framework for
atom-resolved structural analysis and interactive visualization of
twisted bilayer graphene (TBG).
(a) Atom-resolved structure of the two graphene layers with a relative
twist angle $\theta$.
(b)  GPU computational workflow from atomic-coordinate generation
to atomic-site displacement, periodic registry reduction,
AA/AB/BA classification, connected AA-domain identification,
selection of a representative central AA domain, and extraction of the six lattice-neighbor distances used to determine \(L_{\min}\) and \(L_{\mathrm{ave}}\).
(c) NVIDIA Omniverse interface integrating atom-resolved visualization
with AA-domain identification and quantitative moir\'e descriptors,
including $L_{\min}$ and $L_{\mathrm{ave}}$. 
The dashed box marks the interface region whose text is reproduced separately at an enlarged scale for readability.
}
\end{figure*}


\subsection{CPU--GPU benchmark protocol}

The computational benchmark evaluates the calculation triggered by
changing the twist angle from $0^\circ$ to $1^\circ$ using identical
NVIDIA Warp implementations on CPU and GPU backends.

The measured interval includes the structural and moir\'e calculations
performed by the core calculation routine. Initialization, memory
allocation, heatmap generation, user-interface updates, and rendering
are excluded from the timing measurement.

The CPU calculation uses the NVIDIA Warp CPU backend with one thread, while the GPU calculation is performed on an NVIDIA RTX 5090. For every system size and device, 10 warm-up runs are followed by 50 measured repetitions. GPU execution is explicitly synchronized for each timed calculation so that asynchronous CUDA execution is included in the reported latency. Timing is stopped before benchmark data are written to disk; therefore, CSV file I/O does not contribute to the measured calculation time.

The reported latency is the median of the 50 measured runs, and the P10--P90 range is used to characterize run-to-run variation.

The GPU speedup is defined as

\begin{equation}
S(N)
=
\frac{
t_{\mathrm{CPU}}^{\mathrm{median}}(N)
}{
t_{\mathrm{GPU}}^{\mathrm{median}}(N)
}.
\end{equation}

Benchmarks were performed using an AMD Ryzen 9 9950X3D CPU and an NVIDIA RTX 5090 GPU. 
The software environment consisted of Ubuntu 22.04, Python 3.10.12, NVIDIA driver 580.178.04, NVIDIA Warp 1.13.0, and NVIDIA Isaac Sim 6.0.1.


\begin{figure*}
\centering
\includegraphics[width=2\columnwidth]{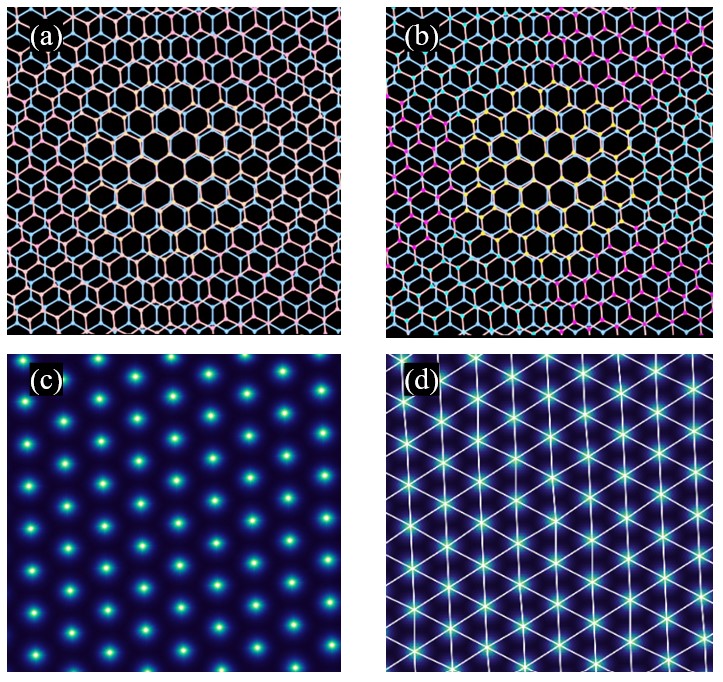}
\caption{\label{fig:rt} 
Atom-resolved registry analysis and local moir\'e-length extraction
for twisted bilayer graphene.
(a) Representative atomic configuration generated at a twist angle
$\theta=5^\circ$.
(b) Atomic-site AA/AB/BA stacking classification obtained from the minimum periodic registry distances, with AA, AB, and BA sites shown in yellow, cyan, and magenta, respectively.
(c) Continuous \(d_{\rm AA}\) field reconstructed from the atom-resolved registry values and projected onto the reference-plane coordinates for visualization.
(d) Representative central AA-domain center and its six lattice-neighbor AA domains. The six center-to-center distances $L_1,\ldots,L_6$ are used to determine the minimum and mean moir\'e length scales, $L_{\min}$ and $L_{\mathrm{ave}}$.
Panels (a,b) show the rotated atomic configuration, whereas panels (c,d) are displayed in the reference-plane coordinates used for structural analysis.
}
\end{figure*}


\begin{figure*}
\centering
\includegraphics[width=2\columnwidth]{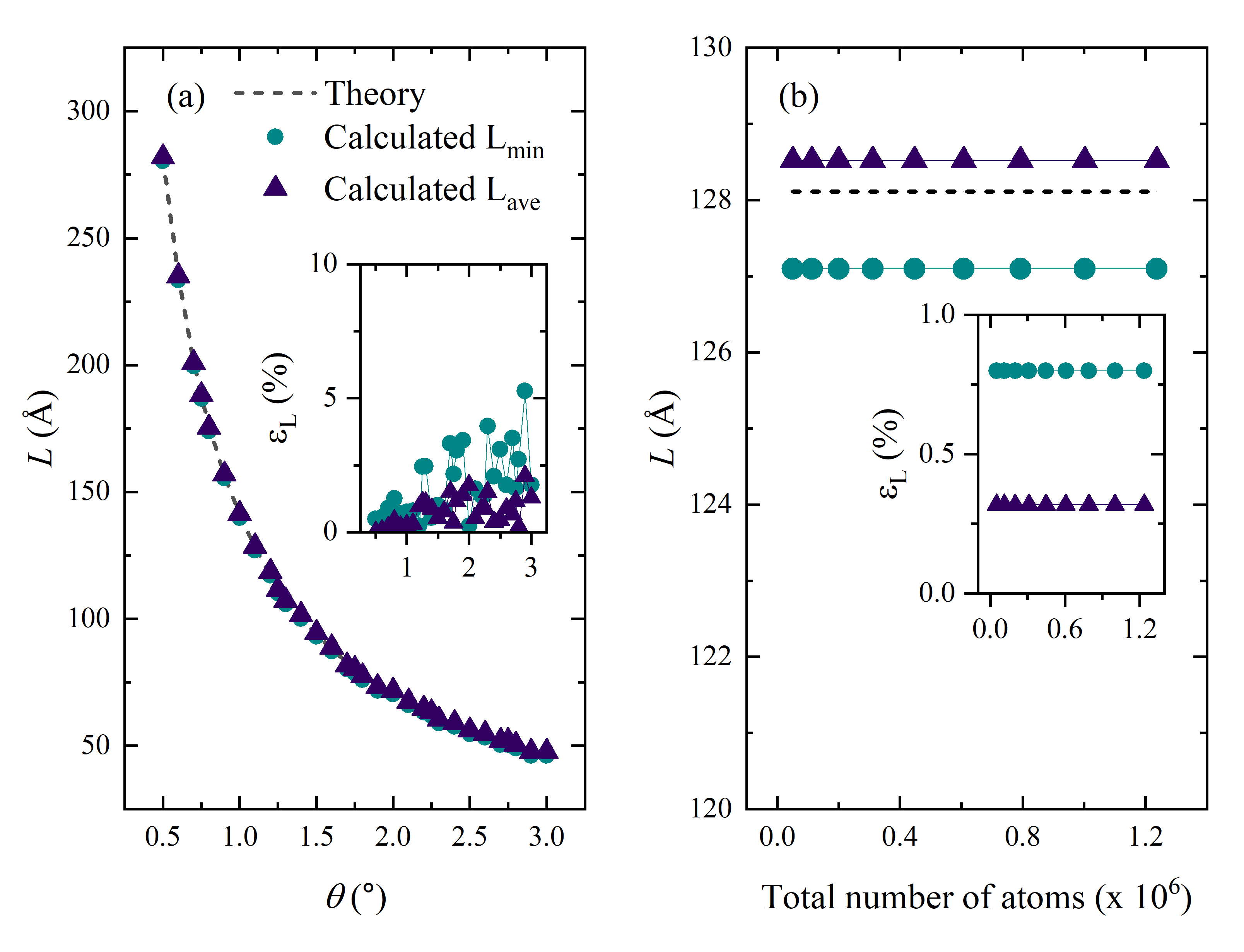}
\caption{\label{fig:rt} 
Geometric consistency of the extracted moir\'e length scales.
(a) Calculated $L_{\min}$ and $L_{\mathrm{ave}}$ as functions of twist angle over $0.5^\circ \leq \theta \leq 3.0^\circ$, compared with the analytical geometric period
$L_{\mathrm{theory}}=a/[2\sin(\theta/2)]$.
(b) Dependence of \(L_{\min}\) and \(L_{\mathrm{ave}}\) on total number of atoms at \(\theta=1.1^\circ\). Both quantities remain nearly constant with increasing system size,
showing weak finite-size dependence over the investigated range.
}
\end{figure*}

\begin{figure*}
\centering
\includegraphics[width=2\columnwidth]{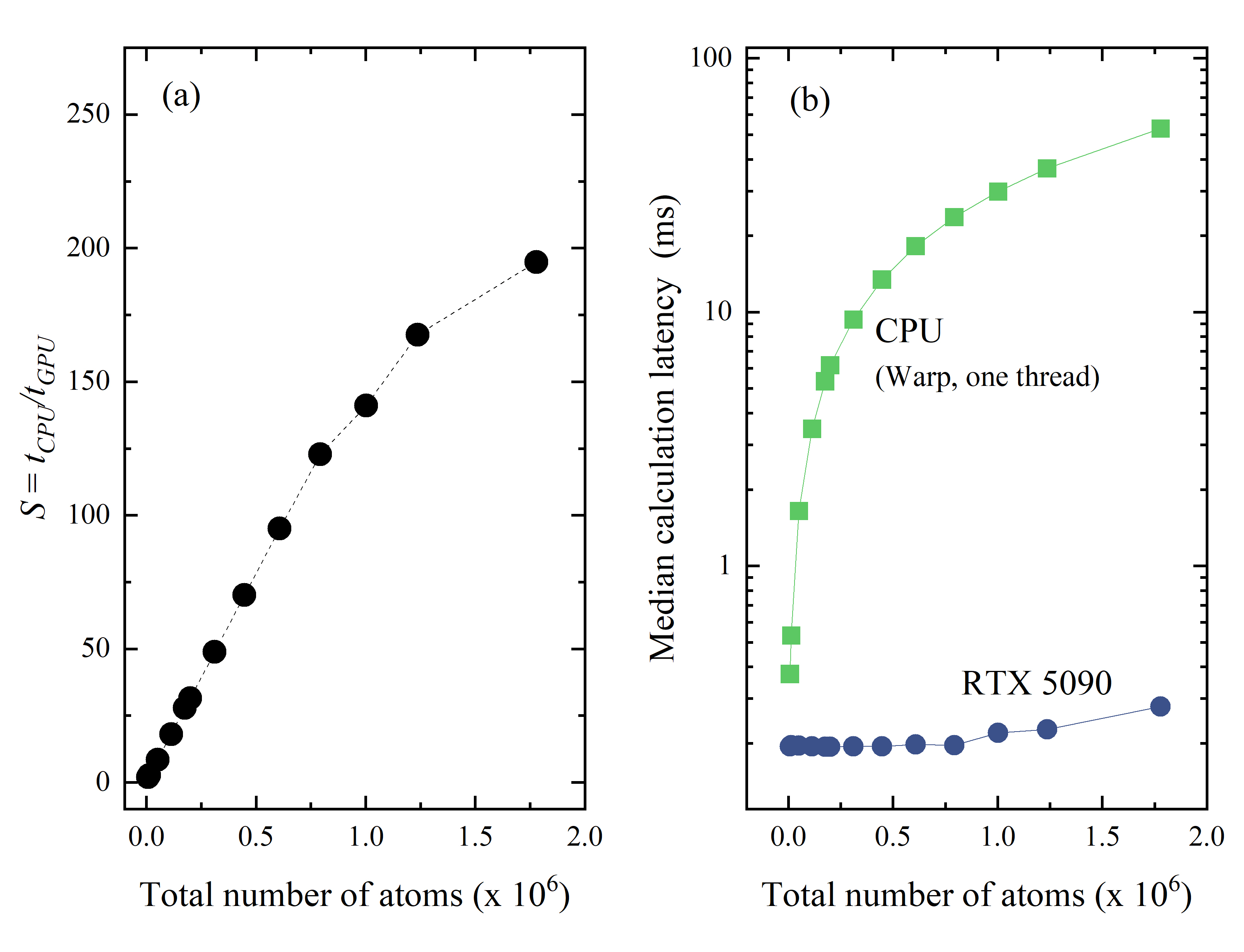}
\caption{\label{fig:speed} 
CPU--GPU performance benchmark for the twist-angle update calculation.
(a) GPU speedup,
$S=t_{\mathrm{CPU}}^{\mathrm{median}}/
t_{\mathrm{GPU}}^{\mathrm{median}}$,
as a function of the total number of atoms in both graphene layers.
(b) Median calculation latency for the single-threaded NVIDIA Warp
CPU backend and NVIDIA RTX~5090 GPU.
For the largest system containing 1,778,700 atoms, the CPU and GPU
latencies are 52.718~ms and 0.2706~ms, respectively, corresponding
to a speedup of 194.8$\times$.
Each point represents the median of 50 measured runs following
10 warm-up iterations; GPU execution is synchronized before and
after each timed call. Initialization, heatmap generation,
user-interface updates, and rendering are excluded.
}
\end{figure*}

\section{Results}

\subsection{Integrated atom-resolved structural-analysis workflow}

Figure~1 summarizes the computational architecture developed in this
work. Starting from the graphene primitive lattice and a prescribed
twist angle $\theta$, explicit coordinates are generated for the
bottom and top graphene layers [Fig.~1(a)]. A variation of $\theta$
changes the atomic configuration directly rather than modifying only
a graphical representation of the moir\'e pattern.

The updated coordinates are subsequently used to evaluate the
atomic-site relative displacement
$\mathbf{u}^{\mathrm{raw}}_i$.
After reduction modulo the graphene primitive lattice, the periodic
distances to the reference AA, AB, and BA registries are evaluated
and used to assign the local stacking configuration.

The AA-classified sites are subsequently grouped into connected
domains using the graphene bond topology. One representative center
is assigned to each domain.

For moir\'e-period extraction, the AA center closest to the geometric
center of the finite graphene flake is selected as a reference.
The distances to its six lattice-neighbor AA domains define the local moir\'e lengths \(L_k\), from which \(L_{\min}\) and \(L_{\mathrm{ave}}\) are obtained [Fig. 1(b)].

The resulting structural data are transferred to the visualization
layer of the workflow, where the individual carbon sites and
registry-derived quantities are represented within NVIDIA Omniverse
[Fig.~1(c)]. Importantly, the visualization is driven by the same
atomic state used for the quantitative structural analysis rather
than by a separately generated moir\'e image.

Quantitative outputs, including $L_{\min}$, $L_{\mathrm{ave}}$,
and computational-performance metrics, are evaluated within the
same workflow.
The resulting architecture therefore links a user-controlled structural parameter to atom-resolved computation, local analysis, quantitative characterization, and three-dimensional visualization within a common interactive loop.

\subsection{Atom-resolved registry and AA-domain identification}

Figure~2 shows the successive structural quantities obtained from the
GPU-based registry analysis. The discrete classification in
Fig.~2(b) identifies AA-, AB-, and BA-like atomic sites from the
minimum periodic registry distance.

The corresponding $d_{\mathrm{AA}}$ values retain continuous
information about the proximity of each atomic site to ideal AA
registry. For visualization, these values are projected onto a
two-dimensional spatial grid, producing the continuous registry map
shown in Fig.~2(c).

Individual AA-like atomic sites do not directly define independent
moir\'e centers because multiple neighboring sites can belong to the
same AA domain. We therefore group bond-connected AA-like sites into
connected components and assign one representative center to each
domain using the minimum-$d_{\mathrm{AA}}$ site.
The resulting AA centers are shown in Fig.~2(d).

The resulting AA-domain centers reproduce the expected moir\'e-scale
spatial arrangement. The AA center closest to the geometric center
of the sample is selected as the reference center, and its six nearest-neighbor AA domains on the indexed triangular moir\'e lattice are identified [Fig.~2(d)].

For the ideal rigid structure, these six distances provide equivalent
local estimates of the moir\'e period within the numerical resolution
of the atom-resolved representation.

\subsection{Geometric consistency of the extracted moir\'e lengths}

Figure~3(a) compares the calculated $L_{\min}$ and $L_{\mathrm{ave}}$ with the analytical geometric period over
$0.5^\circ \leq \theta \leq 3.0^\circ$.
The twist-angle sweep was performed using a fixed shell radius of
$s=320$. Values of $L_{\min}$ and $L_{\mathrm{ave}}$ were retained only when all six valid lattice-neighbor AA domains were identified for the central reference domain.
Both quantities reproduce the expected increase in moir\'e length with decreasing twist angle.

The maximum absolute relative errors over the investigated range are
5.00\% for $L_{\min}$ and 2.06\% for $L_{\mathrm{ave}}$.
The smaller deviation of $L_{\mathrm{ave}}$ reflects the averaging
over the six lattice-neighbor AA domains.

Figure~3(b) examines the dependence on finite system size at
$\theta=1.1^\circ$. Both $L_{\min}$ and $L_{\mathrm{ave}}$ remain
nearly unchanged over the investigated atom-count range, indicating
that the extracted length scale is insensitive to further increases
in system size once the relevant neighboring AA domains are contained
within the finite flake.


\subsection{CPU--GPU acceleration of twist-angle updates}

Figure~4 compares the execution time of the $0^\circ$--$1^\circ$
twist-angle update calculation using the NVIDIA Warp CPU and GPU
backends.

Across the investigated system-size range, the GPU calculation time
varies only weakly with atom count, whereas the CPU latency increases
substantially as the system becomes larger. Consequently, the GPU
speedup increases systematically with system size.

For the largest system investigated, containing 1,778,700 atoms in the two graphene layers, the median CPU and GPU calculation times are 52.718 ms and 0.2706 ms, respectively. The corresponding P10--P90 ranges are 52.611--52.831 ms for the CPU and 0.2628--0.2794 ms for the GPU. This corresponds to a GPU speedup of $S$=194.8.

The benchmark isolates the structural and moir\'e calculation itself;
initialization, heatmap generation, user-interface updates, and
rendering are not included. Therefore, these measurements quantify
the acceleration of the computational core rather than the complete
interactive application latency.

\section{Discussion}

From a computational perspective, the benchmark demonstrates that the primary advantage of GPU execution emerges during repeated angle-dependent structural calculations. The single-threaded CPU latency grows substantially with atom count, whereas the RTX 5090 GPU calculation remains below 1 ms throughout the investigated range, reaching a 194.8$\times$ speedup at 1,778,700 total atoms.
The reported speedup is referenced to the single-threaded NVIDIA Warp CPU backend and therefore characterizes backend-level acceleration rather than a comparison with an independently optimized multicore CPU implementation.
The present timing results characterize the computational core and should not be interpreted as complete application-response latency, because heatmap generation, user-interface updates, and rendering are excluded. NVIDIA Omniverse is used here as an interactive
visualization layer, while a detailed end-to-end visualization
benchmark is left for future work.

The comparison of $L_{\min}$ and $L_{\mathrm{ave}}$ with the
analytical geometrical period provides a direct consistency check
for the registry-based moir\'e characterization. Over the investigated
range of $0.5^\circ$--$3.0^\circ$, the extracted length scales follow
the expected angular dependence, with maximum absolute relative
errors of 5.00\% and 2.06\%, respectively.
The smaller maximum deviation of $L_{\mathrm{ave}}$ is consistent
with the expected reduction in site-discretization sensitivity
obtained by averaging over the six lattice-neighbor AA-domain separations.

\section{Conclusions}

We have developed a GPU-native atom-resolved framework for structural
analysis of rigid twisted bilayer graphene using NVIDIA Warp, coupled
to NVIDIA Omniverse for interactive visualization.
The framework evaluates local AA-, AB-, and BA-like registry directly
from GPU-generated atomic coordinates and groups connected AA-like
sites into representative domains. The minimum and mean center-to-center distances between a representative central AA domain and its six lattice-neighbor AA domains, $L_{\min}$ and $L_{\mathrm{ave}}$, provide
compact numerical measures of the moir\'e length scale.

Over the investigated twist-angle range of
$0.5^\circ$--$3.0^\circ$, the calculated lengths reproduce the
expected analytical angular dependence, with maximum absolute relative
errors of 5.00\% for $L_{\min}$ and 2.06\% for $L_{\mathrm{ave}}$.
The GPU implementation also provides substantial acceleration for
repeated twist-angle calculations. For 1,778,700 atoms in the two
graphene layers, the calculation time is reduced from 52.718~ms on
the single-threaded NVIDIA Warp CPU backend to 0.2706~ms on an
RTX~5090 GPU, corresponding to a speedup of 194.8$\times$.

These results establish a scalable structural-analysis layer for
large atom-resolved moir\'e systems. Future extensions can incorporate
relaxed or defective atomic configurations and couple the same
coordinate representation to real-space tight-binding and other
electronic-structure calculations.

\section*{Code and data availability}

The source code associated with the reproducibility package is currently undergoing institutional and collaborative review for public release. 


\section*{Acknowledgements} \label{sec:acknowledgements}
    The authors acknowledge the institutional support of ALL.AI Inc. and the University of Florida.

\bibliography{main.bib}

\end{document}